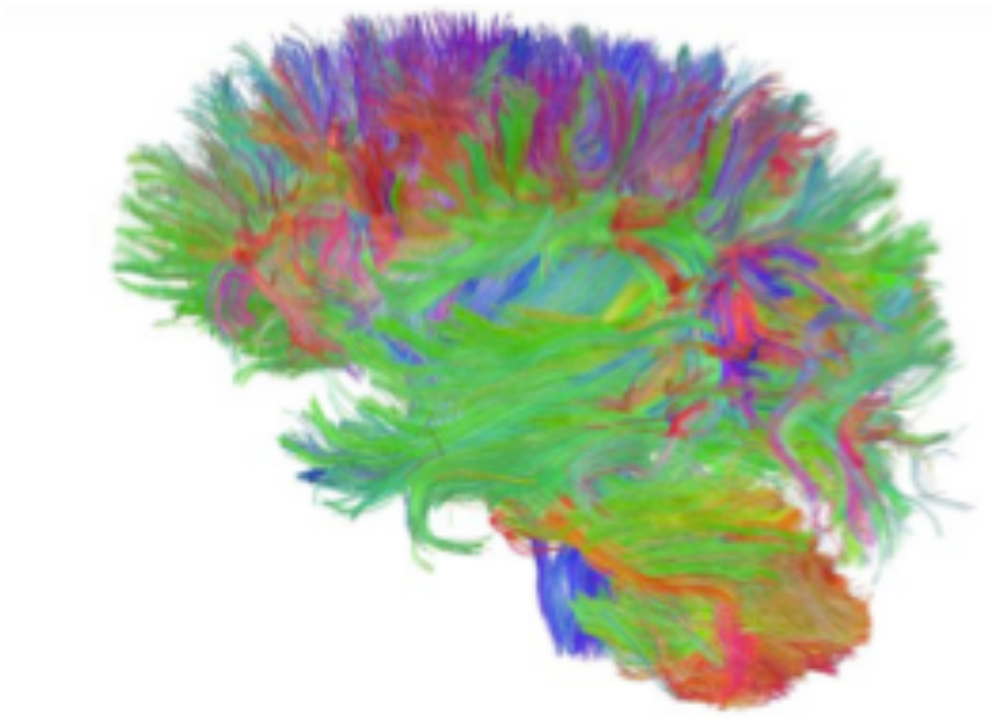

Marcus Kaiser

# A hierarchical network organization helps to retain comparable oscillation patterns in rats and human-sized brains

# A hierarchical network organization helps to retain comparable oscillation patterns in rats and human-sized brains

Marcus Kaiser[1,2,3*]

1 School of Computing Science, Newcastle University, United Kingdom
2 Institute of Neuroscience, Newcastle University, United Kingdom
3 Department of Brain and Cognitive Sciences, Seoul National University, Korea

**Correspondence:**
Dr Marcus Kaiser
School of Computing Science
Newcastle University
Claremont Tower
Newcastle upon Tyne, NE1 7RU
United Kingdom
m.kaiser@ncl.ac.uk

**Summary:**
Activity in coupled systems is often oscillatory, for example, the firing pattern of neuronal populations. Whereas these oscillations have been studied predominantly in local circuits, here we show how the topology of large-scale networks, leading to large feedback loops, influences oscillations in the resting state. We find that the hierarchical modular organization of neuronal networks supports distinct spectral patterns of neural rhythms similar to those observed experimentally in different species such as rat and human. For individual neurons, multiple peak frequencies with non-integer ratios between subsequent peaks occurred. These ratios occurred both for models with the spatial size of the rat as well as for the human brain. We argue that the potential influence of longer connections, and thus longer delays, are balanced by a reduced number of long-distance connections in larger brain networks. In conclusion, we show that a hierarchical neuronal network provides a scalable backbone for multiple brain rhythms. This structural backbone could complement well-studied regional and cellular mechanisms for the generation or prevention of rhythms.

## 1. Introduction

Feedback loops play an important role in complex neural systems. However, many studies focus on direct neural feedback between two entities, such as between local inhibitory and excitatory populations (Buzsaki et al., 2004), rather than large-scale networks between brain regions (Izhikevich and Edelman, 2008a). In addition, large-scale models often use mean-field models (Deco et al., 2009), aggregating a population of neurons, rather than having individual neurons as network nodes (Deco et al., 2008). In this study, we observe a large-scale model of individual neurons to evaluate the role of spatial distances, and therefore delays, on the frequency patterns of the whole brain and of individual nerve cells. We are providing models based on different brain sizes, and consequently delays, in rat and human-sized cortical networks. Here, a similar size does not mean the same number of neurons but the same spatial distance between neurons that also informs the delays for signal propagation.

Recurrent loops and oscillations are integral components of brain function (Singer and Gray, 1995; van Rossum et al., 2008). Oscillations are a near-ubiquitous feature of brain networks, first discovered using electroencephalography (EEG) (Berger, 1929). Distinct frequency bands for oscillations (e.g. theta, alpha, beta, or gamma) are comparable for different species (Buzsaki, 2006) and similar frequency bands are observed across different levels of neural organization–from EEG signals to local field potentials. Recent reports have also demonstrated the existence of multiple oscillation frequencies, associated with whole brain recordings, in small isolated sections of cortex *in vitro* (Roopun et al., 2008, 2008b), with ratios between neighboring peaks in the frequency distribution being close to the golden ratio $\varphi$ (1.618) or $\varphi^2$ (2.618) (see Figure 3 and Figure 6 in Roopun et al., 2008a as an example). The latter value was close to the *in vivo* ratio between mean oscillation frequencies of $e$ (2.718) (Buzsaki, 2006). EEG frequencies in humans were also suggested to be arranged as harmonics around the golden ratio (Weiss and Weiss, 2003). Such irrational ratios of frequency peaks are beneficial for the brain as two connected populations with different peaks will show minimal interference and can therefore process information independently (Roopun et al., 2008a). In other terms, these ratios allow 'channels' to coexist with minimal temporal interference (multiplexing). Indeed, it has been proposed that sensory information is processed more efficiently if multiple 'channels' with differing temporal scales are used (Rubinov et al., 2009). Are these frequency patterns linked to the topological and spatial organization of neuronal networks or can they only be explained through cellular properties?

Computational models and experimental studies on oscillations have usually focused on specific local circuits as rhythm generators, e.g. thalamo-cortical (Destexhe et al., 1998) or hippocampal loops (Dyhrfjeld-Johnsen et al., 2007; Traub et al., 2005; Traub et al., 1999). An example for local mechanisms is the role of inhibitory interneurons in the generation of gamma or fast rhythms in cortical tissue (Traub et al., 1996; Wendling et al., 2002; Bartos et al., 2007). Extending to larger networks, previous modeling studies looked at the role of direct conduction delays on synchronization (Kopell et al., 2000), firing-rate neuronal models (Roxin et al., 2005), of coupling strength on cortical rhythms (Anastassiou et al., 2011; David and Friston, 2003; David et al., 2005; Jansen and Rit, 1995), and of the mean path length between neurons as generators and determinants of characteristic oscillation frequencies (Traub et al., 1999).

We hereby investigate the role of hierarchy and brain size on the potential oscillation patterns of neural systems. We present a conceptual model of why larger brains (e.g. for human) can show similar frequency patterns compared to smaller brains (e.g. for rodents).

## 2. Materials and Methods

Calculations were performed on a 48-core AMD Opteron Linux cluster using Matlab R2009a (Mathworks Inc., Natick, MA). Scripts are available at http://www.biological-networks.org.

### 2.1. Network Model

In this study, we simulate large-scale hierarchical networks from brain connectivity available for different species like human (Hagmann et al., 2008) and rat (Burns and Young, 2000). At the highest level of the hierarchy, the networks consist of regions of interest (ROI) in humans and cortical and sub-cortical areas in rats, with connections based on diffusion spectrum imaging or tract tracing, respectively. At the intermediate level, each area or ROI includes several columns, which at the lowest level contain individual excitatory or inhibitory neurons (see Figure 1). The connectivity within and between different levels as well as the delays for signal propagation is based on the neuroanatomy of both species (Douglas and Martin, 2007; Thomson and Lamy, 2007; Young, 2000).

The rat fiber-tract connectivity between regions consisted of the hippocampus and the limbic cortex (Burns and Young, 2000), therefore including cortical as well as a few sub-cortical structures (see Figure S1 in the supplemental material). For simplicity, all regions were treated as cortical with an internal modular architecture that we called columns. However, it is known that also sub-cortical structures–though not containing columns–exhibit a modular architecture of local clustering. For example, the hippocampus shows properties of small-world networks in that direct neighbors of a neuron, to which that neuron is connected, are more often connected between themselves than would be expected for a random homogeneous organization (Buzsaki et al., 2004; Netoff et al., 2004).

At the lower level of the hierarchical network, containing individual neurons as nodes, connections between neurons in two regions were established when an existing fiber-tract was reported in the literature. The literature on the rat also included information on the strength of connections (weak; medium; strong; unknown). However, using these ordinal values would have obscured the calculation, as it was not reported whether a strong connection is 2 or 10 times as strong as a medium one; the treatment of connections with unknown strength would have been even more arbitrary. Also for the human network, we not include information on fiber strength as we use regions of interest of comparable size as high-level nodes. Therefore, all known fiber tracts were assumed to have the same strength. Note, that this assumption does not change the frequency peaks: increasing the number of connections between two areas would increase the number of paths with a certain length (signal amplitude) but not the path lengths (feedback delay and corresponding frequency).

Networks were binary (1: existing, 0: absent connection). Analogue to the rat (*rattus norvegicus*), the global level consisted of 23 regions where connectivity was based on known anatomical connections (Burns and Young, 2000). The resulting fiber-tract network between areas had an edge density of 41.5%, a clustering coefficient (Watts and Strogatz, 1998) of 52.0%, and a characteristic path length of 1.7. Extending the network model, regions consisted of 25 columns with 10 neurons each (see section 2.2 for network properties). Connections between neurons were randomly and independently established with probabilities of 16% within columns, of 8% between columns of the same region, and with 4% between columns of different regions if a connecting fiber-tract exists. These values were in the range of those reported in the literature (Douglas and Martin, 2007; Thomson and Lamy, 2007; Young, 2000). A recent study shows that the degree distribution has an important role in the stability of the oscillation modes (Roxin 2011). Connections are randomly and independently drawn; our network constitutes an Erdős–Rényi graph with binomially distributed in- and out-degrees.

The human connectivity provided in (Hagmann et al., 2008) consisted of 998 ROIs covering the n=66 anatomical regions of cortices of both hemispheres but excluding subcortical regions (see Figure S2 in the supplemental material). In the next level of hierarchy, each region consists of 2 columns with 10 neurons each. The connections between neurons were established following the same probabilities as in the rat case.

## 2.2 Connectivity characteristics

*Distances and average fiber lengths*

For the *human*, the average distance between two nodes was based on the surface coordinates of the right hemisphere based on data of the CARET software (http://brainvis.wustl.edu/wiki/index.php/Caret:About). The three-dimensional Euclidean distance between all pairs of 1,000 randomly chosen surface coordinates was calculated. The average distance was 66.6 mm – 39% of the maximal distance (168 mm).

For the *rat*, no surface coordinates existed and we used a spherical surface as an approximation. This is possible as the brain is lissencephalic so that convolutions do not influence the calculation. It is known that the average Euclidean distance between two nodes on a sphere is 67.5% of the maximum distance, which corresponds to the diameter of the sphere (note, that the average distance for lissencephalic brains is much higher than for convoluted cortical surfaces). Taking the length of a hemisphere (12.5 mm, see rat atlas at http://www.loni.ucla.edu) as the maximum distance, the average distance is 8.4 mm.

Note that the previous numbers only show the average fiber length for all-to-all connectivity. For neural systems, only a fraction of all possible connections exists. In addition, the probability that two neurons are connected decays with the distance between two nodes. Therefore, the average distance will be much lower than shown above. The schematic calculation (Figure 7a) would indicate average fiber lengths of 5.1 mm, 5.5 mm, 5.5 mm, and 5.5 mm (rat, cat, macaque, and human, respectively).

Using a variant of Floyd's algorithm (Cormen et al., 2009), we calculate not only the length of the shortest path between any two nodes but also how many paths of such a minimal length exist. The length of a path (number of intermediate connections) from node A to node B and back to node A determines the length of a feedback loop.

*Network properties*

The network of the rat, consists of $N_e$=5750 excitatory neurons had an edge density of 0.48% (ratio between the number of existing edges and the number of all possibly existing edges), a clustering coefficient (Watts and Strogatz, 1998) of 4.68%, and a characteristic path length of 3.3. Comparable random networks had a clustering coefficient of 0.48% and a characteristic path length of 2.9, indicating that the hierarchical network is a small-world network (Watts and Strogatz, 1998). The human brain network consists of $N_e$=19960 excitatory neurons with an edge density of 0.81%, a clustering coefficient of 10.2% and a characteristic path length of 5.8, exhibiting small-world features. The connectivity pattern between neurons is drawn following the same connection probability as for the rat network detailed above.

In the network for both species, we add 20% of the total number of nodes acting as inhibitory neurons with a connectivity pattern dependent on the distance between the neurons, avoiding long-range inhibitory connections.

*Scaling of edge density with brain size*

Brain size is approximated by the maximum distance within the right hemisphere (sagittal plane). Maximum distances, taken from www.brainmuseum.org, are 15.4 mm (rat; *rattus norvegicus*), 40 mm (cat; *felis sylvestris*), 54.7 mm (macaque; *macaca mulatta*), and 152.5 mm (human; *homo sapiens*). The edge density *d* defines the proportion of existing cortico-cortical fiber tracts out of all possibly existing fiber tracts and is calculated using the number of fiber tracts *E* and the number of regions *N* by $d = E / (N * (N-1))$. Edge densities for the above named species are 41.5% (rat (Burns and Young, 2000)), 30% (cat (Hilgetag et al., 2000)), 26.9% (macaque (Kaiser and Hilgetag, 2006)), and 10% (human based on diffusion tensor imaging).

## 3. Results

### 3.1 Difference between regional and global scale

We have seen that multiple frequency peaks can occur at the global scale. One characteristic of a scale is the length of potential feedback loops in the system, i.e. the number of synapses that are on the pathway from one neuron, including several intermediate neurons, back to itself. Therefore, the length of a feedback loop neither signifies spatial distance nor time delays but the length of a path from one node back to itself (that is, a cycle in graph theory terms (Sporns et al., 2000)). The length of that path is the number of connections that are part of the path.

As shown in Figure 3A for the rat connectivity, the global network contains a higher proportion of long feedback loops. The distribution of the regional network, within one region, shows a remarkably low overlap with the global distribution containing a higher proportion of short feedback loops whereas the overlap is larger between the global and a comparable random network (Figure 3B). Note that while the random network shows a more uniform connection pattern, our hierarchical networks show a higher edge density within regions than for the network as a whole. This might indicate that some loop delays, and therefore some frequencies, are more likely to occur on the global rather than the regional scale and vice versa.

We also observe similar features for the human-sized network. As shown in Figure 3C, the distribution of loops for the regional network shows a sharp peak at a loop length of size 5. This is in contrast with the wider distribution for the global network that is centered at a larger loop length. We also observe a low overlap between both distributions suggesting that certain frequencies tend to occur at a global rather than at a regional level and vice versa. Again, for the global level (Figure 3D) loops for the actual network were longer than for a random network.

In summary, feedback loops were short within areas or regions of interest supporting high frequency oscillations *in vivo* and for resected tissue. At the same time, the complete network showed longer loops, even longer than for random networks, potentially supporting the formation of low-frequency oscillations.

### 3.2 Scaling of potential frequencies with brain size of different species

We have seen that a model of brain rhythms, which includes delays and a hierarchical network organization, can generate distinct frequency peaks. What happens if the average distance, and therefore the average delays, increases for a larger brain? If average fiber lengths increase with brain size, we should expect that different potential oscillations are supported. However, experimental recordings show that frequency bands in the relatively small rat brain are similar to the ones in the larger human brains or other mammalian species. How then can the pattern of delays, and thus preferred frequencies, stay comparable when the size of the brain increases by an order of magnitude, from about 10 mm in rats to more than 150 mm in humans?

It is known that the probability that two neurons are connected decays almost exponentially with distance (Hellwig, 2000; Kaiser et al., 2009; Schüz et al., 2005): connections over a long-distance are less likely than short-distance connections. Therefore, comparing networks of different sizes, only few of the long-distance connections which could theoretically be established for larger brains actually exist. How would this affect the average connection length? In different organisms, ranging from neuronal connectivity in *C. elegans* and layers in the rat visual cortex to fiber-tract connectivity in the macaque, the actual distribution could best be approximated by a Gamma probability density function (Kaiser et al., 2009). We therefore approximate the distance-dependent connection probability by a Gamma function (shape parameters a=1.8 and scale parameter b=3). The distribution for different brain sizes (insets of Figure 4A) depended on the maximal distance in each brain. Based on these distributions, we calculated the average distance of a successfully established connection (green vertical lines in Figure 4A). These average distances were almost identical despite differences in brain size with average distances of 5.1 mm, 5.5 mm, 5.5 mm, and 5.5 mm for rat, cat, macaque, and human, respectively. Whereas the actual Gamma distributions at the global level might differ between species (there is no experimental data of connection lengths of complete connectomes except for *C. elegans*), drastic changes are needed to yield highly different average distances. In other words, the average distances across species are similar due to the similar distance dependence of the likelihood of connections between two neurons. Therefore, we suggest that average connection lengths, and therefore average delays given comparable myelination, are similar across brain sizes enabling comparable frequency distributions at the global level.

As a consequence of the Gamma distribution, we would expect relatively fewer long-distance connections, or fiber-tracts, in larger brains. Indeed, the proportion of existing fiber-tracts relative to all possible ones, the edge density, seems to decrease exponentially with brain size (Figure 4B). This relates to the idea that for a low number of neurons in small brains, a larger proportion of potential connections can be established. That means that the average number of synapses per neuron, rather than the edge density of connections, remains constant (Striedter, 2004). Therefore, we expect that changes in brain size have little influence on average delays and frequency oscillations.

## 4. Discussion

We suggest that the similarity between the pattern of frequencies preferred by hierarchical, whole brain networks, and the pattern of frequencies observed in intrinsic cellular and local microcircuit recordings is not circumstantial. Such a hierarchical system provides an ideal substrate for information processing on multiple spatial and temporal scales simultaneously. Indeed, the same tissue might show several populations with distinct frequencies with minimal interference between them ensuring a longer time for stable oscillations (Roopun et al., 2008a, 2008b). In addition, reducing the fiber-tract connectivity between regions as the brain size increases not only preserves existing frequency bands but prevents very low-frequency (VLO) oscillations due to increased delays of long-distance connections: a natural hum-rejection mechanism. Finally, keeping similar ratios across species might provide an evolutionary advantage in that mechanisms developed earlier (e.g. in the rat) can be re-used for phylogenetically more recent species (e.g. human). The described preference for multiple global frequency bands suggests that delay-based oscillations might be a default state of healthy neural systems. It also suggests that long-distance connectivity can modulate oscillation patterns within regions. Therefore, regional and cellular mechanisms might be as important in the *prevention* as well as the generation of oscillations. Future theoretical and experimental studies linking network topology and oscillations are needed to elucidate these questions.

A limitation of the current approach is that detailed information about the cortical columnar organization is not available to the same extent for all cortical areas; e.g. there has historically been a focus on sensory and motor areas with fewer studies on other parts of the cortex (Mountcastle, 1997). In addition, detailed models would also need to take into account characteristics of the observed species (Herculano-Houzel et al., 2008) many of which are not quantitatively described within the neuroanatomical literature (Crick and Jones, 1993).

## Tables

**Table 1.** Default values and ranges of model parameters. Values show the default value while ranges in square brackets show values for control calculations.

| Parameter | Description | Default value |
|---|---|---|
| $N_c$ | Neurons per column | 10 (rat)<br>10 (human) |
| $N_r$ | Columns per region | 25 (rat)<br>2 (human) |
| $N_{regions}$ | Number of regions | 23 (rat)<br>998 (human) |
| N | Total number of neurons<br>($N = N_c\ N_r\ N_{regions}$) | 5,750 (rat)<br>19,960 (human) |

**Figures**

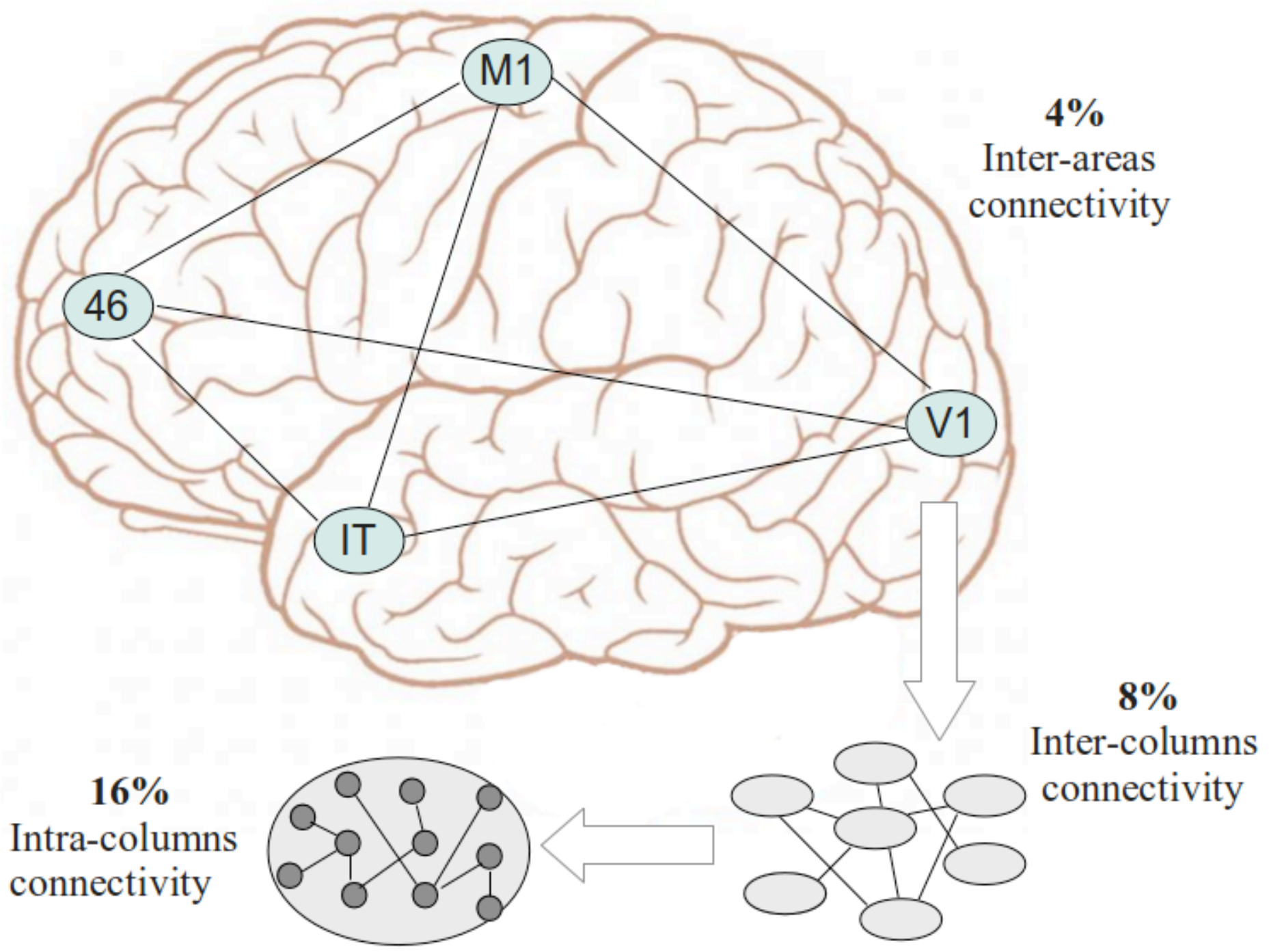

**Figure 1: Schematic of the hierarchical network.** The hierarchical network consist of three different levels: at regional level, connections between neurons of physically connected regions are drawn with a probability of 4%. Within a region, neurons are connected with a probability of 16% (8%) if they are in the same (different) columns. The visual area was used for illustration purposes but the remaining areas follow similar subdivision. At the last level of the hierarchy are the individual neurons, which connectivity is schematically represented.

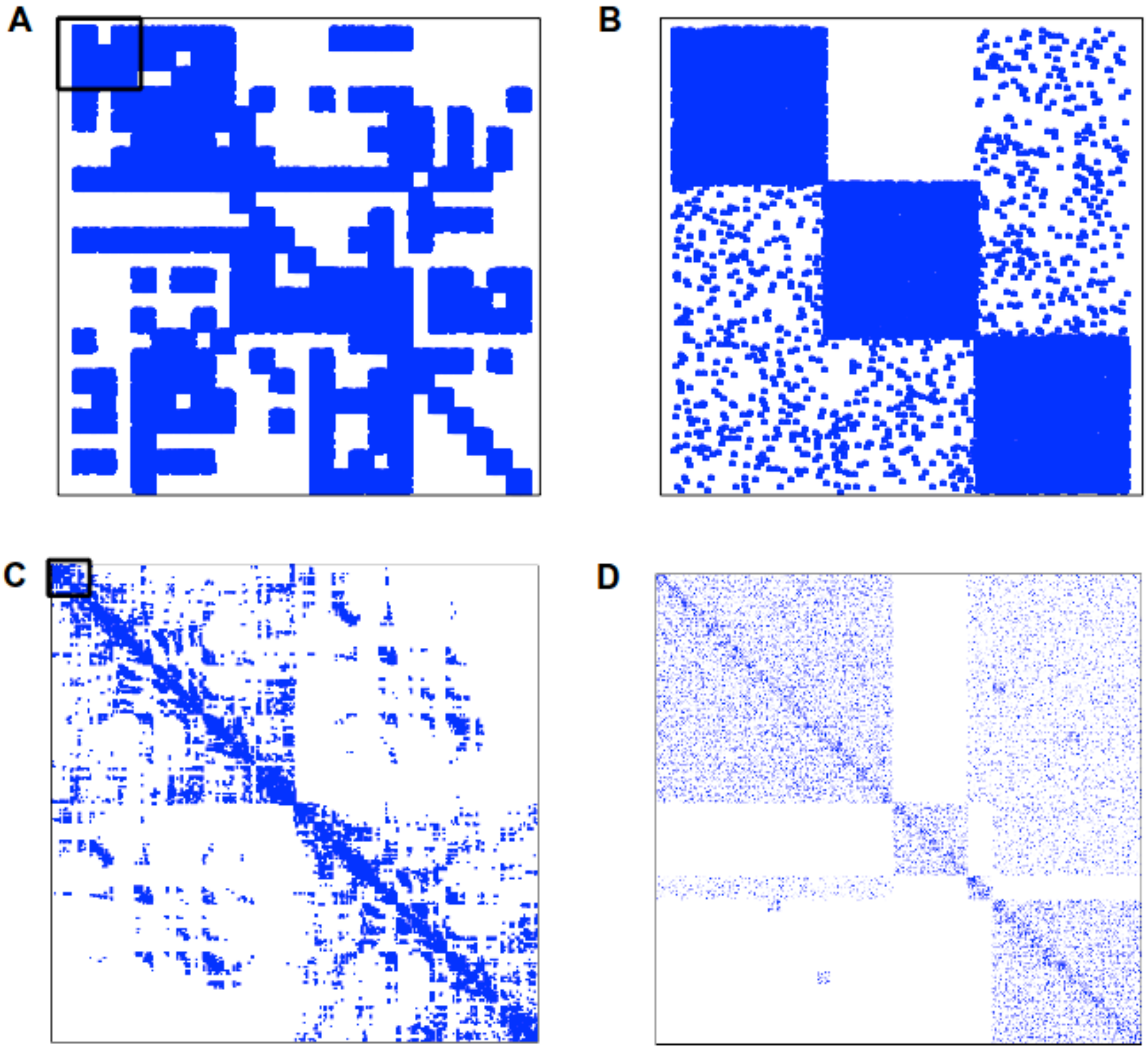

**Figure 2: Adjacency matrix of the global and local network** (dots represent existing connections, rows represent sources and columns targets of projections). (**A**) The global network for the rat connectivity consisted of 23 regions with 250 nodes each. (**B**) Zoom-in on three regions (square in A). Each region formed a local network with 25 columns of 10 neurons each as building blocks. Connections were only established between regions that are connected by a fiber tract; otherwise, connections are absent (white space). (**C**) The global cortical network for the human connectivity consisted of 66 cortical areas of both hemispheres parcellated into 998 ROI containing 2 columns and 10 nodes each. (**D**) Zoom-in on different regions (square in C). Each region formed a local network with 2 columns of 10 neurons each as building blocks.

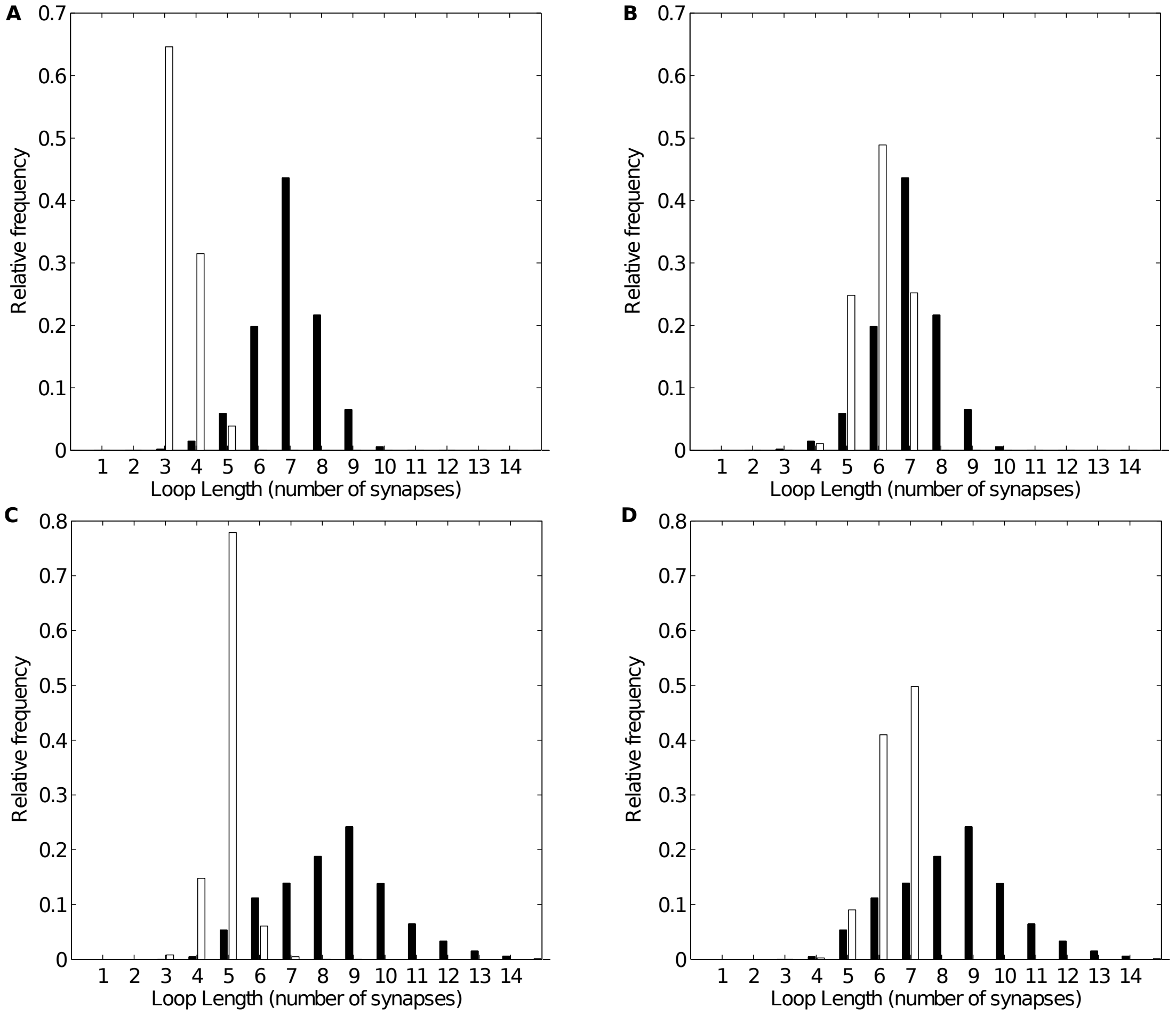

**Figure 3: Distribution of path lengths.** (**A**) Relative frequency of occurrences of a loop with *k* intermediate edges for the local (white) and global (black) hierarchical network in the rat case. (**B**) Distribution for the global hierarchical in the rat case (black) and a random non-hierarchical (white) network. (**C**) Relative frequency of occurrences of a loop with *k* intermediate edges for the local (white) and global (black) hierarchical network in the human connectivity. (**D**) Distribution for the global hierarchical network in the human case (black) and a random non-hierarchical network (white).

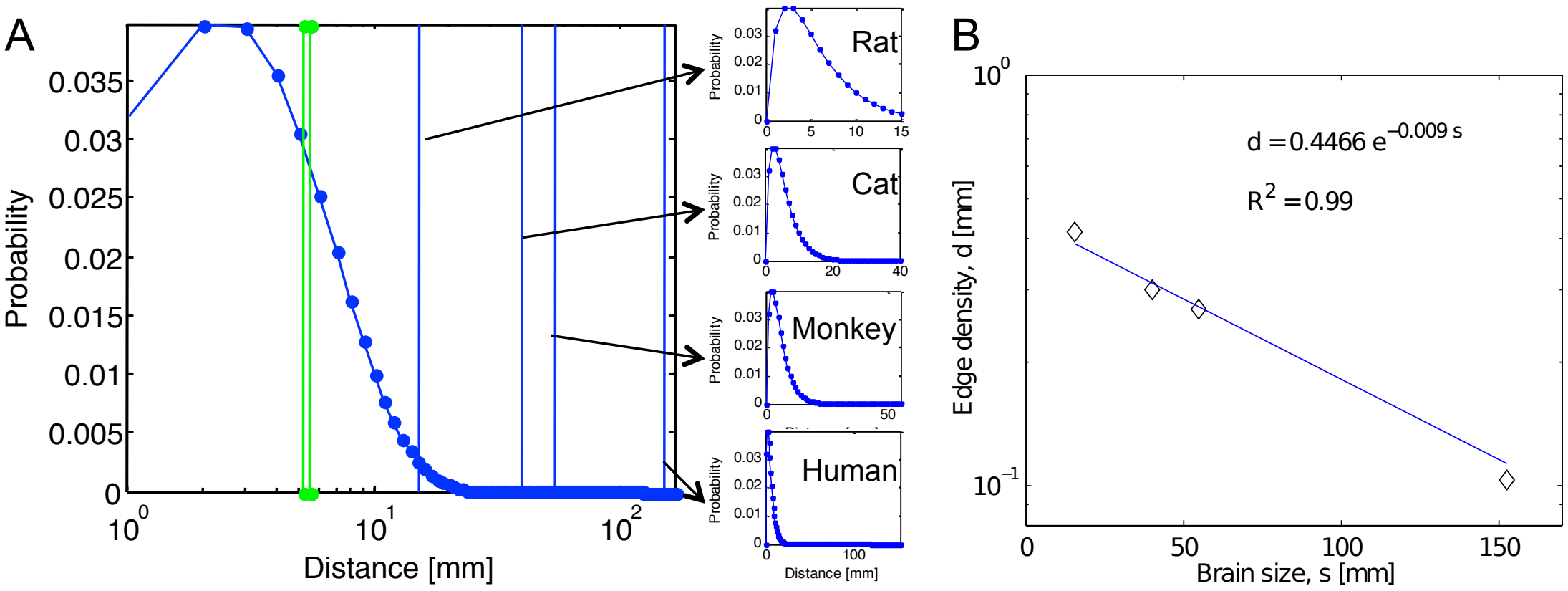

**Figure 4: Scaling of average axon length with brain size (schematic).** (**A**) The probability that two neurons are connected decreases with the distance between two neurons (approximated as Gamma probability density function with shape parameters a=1.8 and scale parameter b=3). The vertical blue lines indicate different sizes of the brain (maximum distance in rat, cat, macaque monkey, and human). Green lines show the average distance for the four species based on the distribution up to the maximum distance (insets to the right). (**B**) Edge density of cortico-cortical fiber tract connectivity decays exponentially with brain size estimated by the maximum distance within one hemisphere (rat, cat, macaque and human). Note, that the edge density for the human is based on diffusion tensor imaging (DTI).

## *Supplementary material*

| | CA1 | CA3 | DG | ENT | PAR | POST | PRE | SUB | LM | MM | SUM | TM | ACA | ILA | PL | PRh | RSP | AD | AM | AV | IAM | LD | MD |
|---|---|---|---|---|---|---|---|---|---|---|---|---|---|---|---|---|---|---|---|---|---|---|---|
| CA1 | 0 | 0 | 1 | 1 | 1 | 1 | 1 | 1 | 0 | 0 | 0 | 0 | 0 | 1 | 1 | 1 | 1 | 0 | 0 | 0 | 0 | 0 | 0 |
| CA3 | 1 | 0 | 1 | 1 | 1 | 0 | 1 | 1 | 0 | 0 | 0 | 0 | 0 | 0 | 0 | 0 | 0 | 0 | 0 | 0 | 0 | 0 | 0 |
| DG | 1 | 1 | 0 | 0 | 0 | 1 | 1 | 1 | 0 | 0 | 0 | 0 | 0 | 0 | 0 | 0 | 0 | 0 | 0 | 0 | 0 | 0 | 0 |
| ENT | 1 | 0 | 1 | 0 | 1 | 1 | 1 | 1 | 0 | 1 | 0 | 0 | 1 | 0 | 1 | 1 | 1 | 0 | 0 | 0 | 0 | 0 | 1 |
| PAR | 1 | 0 | 1 | 1 | 0 | 1 | 1 | 1 | 1 | 0 | 0 | 0 | 0 | 0 | 0 | 0 | 1 | 1 | 0 | 1 | 0 | 0 | 0 |
| POST | 0 | 0 | 0 | 1 | 1 | 0 | 1 | 0 | 1 | 0 | 0 | 0 | 0 | 0 | 0 | 1 | 1 | 1 | 0 | 1 | 0 | 1 | 0 |
| PRE | 0 | 0 | 1 | 1 | 1 | 1 | 0 | 1 | 1 | 1 | 0 | 0 | 0 | 0 | 0 | 0 | 1 | 1 | 0 | 1 | 0 | 1 | 0 |
| SUB | 1 | 1 | 1 | 1 | 1 | 1 | 1 | 0 | 1 | 1 | 1 | 1 | 1 | 1 | 1 | 1 | 1 | 0 | 1 | 1 | 1 | 0 | 0 |
| LM | 0 | 0 | 0 | 0 | 0 | 0 | 0 | 0 | 0 | 0 | 0 | 0 | 0 | 0 | 0 | 0 | 0 | 1 | 0 | 0 | 0 | 0 | 0 |
| MM | 0 | 0 | 0 | 0 | 0 | 0 | 0 | 0 | 0 | 0 | 0 | 0 | 0 | 0 | 0 | 1 | 0 | 1 | 1 | 1 | 1 | 0 | 0 |
| SUM | 1 | 1 | 1 | 1 | 1 | 1 | 1 | 1 | 1 | 1 | 0 | 0 | 0 | 0 | 1 | 1 | 0 | 1 | 0 | 0 | 0 | 0 | 0 |
| TM | 0 | 0 | 0 | 0 | 0 | 0 | 0 | 0 | 1 | 1 | 0 | 0 | 0 | 0 | 0 | 1 | 0 | 0 | 0 | 0 | 0 | 0 | 0 |
| ACA | 0 | 0 | 0 | 1 | 0 | 1 | 1 | 0 | 1 | 1 | 0 | 0 | 0 | 1 | 1 | 1 | 1 | 0 | 1 | 1 | 0 | 1 | 1 |
| ILA | 0 | 0 | 0 | 0 | 0 | 0 | 0 | 0 | 1 | 1 | 1 | 1 | 1 | 0 | 1 | 1 | 1 | 0 | 1 | 1 | 1 | 0 | 1 |
| PL | 0 | 0 | 0 | 1 | 0 | 0 | 1 | 0 | 1 | 1 | 1 | 1 | 1 | 1 | 0 | 1 | 1 | 0 | 1 | 1 | 1 | 1 | 1 |
| PRh | 1 | 0 | 0 | 0 | 1 | 1 | 0 | 1 | 0 | 0 | 0 | 0 | 0 | 0 | 1 | 0 | 1 | 0 | 0 | 0 | 0 | 0 | 0 |
| RSP | 0 | 0 | 0 | 1 | 1 | 1 | 1 | 0 | 0 | 1 | 0 | 0 | 1 | 0 | 0 | 0 | 0 | 1 | 1 | 1 | 0 | 1 | 0 |
| AD | 1 | 1 | 0 | 1 | 1 | 1 | 1 | 0 | 0 | 0 | 1 | 0 | 1 | 0 | 0 | 1 | 1 | 0 | 0 | 0 | 0 | 0 | 0 |
| AM | 0 | 1 | 0 | 1 | 1 | 0 | 1 | 1 | 0 | 0 | 0 | 0 | 1 | 1 | 1 | 1 | 1 | 0 | 0 | 0 | 0 | 0 | 0 |
| AV | 1 | 1 | 0 | 1 | 1 | 1 | 1 | 1 | 0 | 0 | 1 | 0 | 1 | 0 | 0 | 1 | 1 | 0 | 0 | 0 | 0 | 0 | 0 |
| IAM | 0 | 0 | 0 | 1 | 0 | 0 | 0 | 0 | 0 | 0 | 0 | 0 | 1 | 0 | 0 | 1 | 1 | 0 | 0 | 0 | 0 | 0 | 0 |
| LD | 1 | 1 | 0 | 1 | 1 | 1 | 1 | 0 | 0 | 0 | 0 | 0 | 1 | 0 | 1 | 0 | 1 | 0 | 0 | 0 | 0 | 0 | 0 |
| MD | 0 | 0 | 0 | 1 | 0 | 0 | 0 | 0 | 0 | 0 | 0 | 0 | 1 | 1 | 1 | 1 | 1 | 0 | 0 | 0 | 0 | 0 | 0 |

**Figure S1.** Rat inter-regional fiber-tract network. 1: existing fiber-tract; 0: non-existing or non-tested fiber tract. Rows correspond to sources and columns to targets of a projection.

The labels of the matrix represent the following regions:

| | | | |
|---|---|---|---|
| ACA | Anterior cingulate area | MD | Mediodorsal dorsal nucleus of the thalamus |
| AD | Anterodorsal nucleus of the thalamus | MM | Medial mammillary nucleus |
| AM | Anteromedial nucleus of the thalamus | PAR | Parasubiculum |
| AV | Anteroventral nucleus of the thalamus | PL | Prelimbic area |
| CA1 | Ammon's horn, field CA1 | POST | Postsubiculum |
| CA3 | Ammon's horn, field CA3 | PRE | Presubiculum |
| DG | Dentate gyrus | PRh | Perirhinal region |
| ENT | Entorhinal area | RSP | Retrosplenial area |
| IAM | Interoanteromedial nucleus of the thalamus | SUB | Subiculum |
| ILA | Infralimbic area | SUM | Supramammillary nucleus |
| LD | Lateral dorsal nucleus of the thalamus | TM | Tuberomammillary nucleus |
| LM | Lateral mammillary nucleus | | |

| | | | | | | | | | | | | | | | | | | | | | | | | | | | | | | | | | | | | | | | | | | | | | | | | | | | | | | | | | | | | | | | | | | |
|---|---|---|---|---|---|---|---|---|---|---|---|---|---|---|---|---|---|---|---|---|---|---|---|---|---|---|---|---|---|---|---|---|---|---|---|---|---|---|---|---|---|---|---|---|---|---|---|---|---|---|---|---|---|---|---|---|---|---|---|---|---|---|---|---|---|---|
| rBSTS | 0 | 0 | 0 | 0 | 0 | 0 | 1 | 1 | 1 | 0 | 1 | 0 | 0 | 0 | 1 | 0 | 0 | 1 | 0 | 1 | 0 | 0 | 0 | 1 | 0 | 0 | 0 | 0 | 0 | 1 | 1 | 0 | 1 | 0 | 0 | 0 | 0 | 0 | 0 | 0 | 0 | 0 | 0 | 0 | 0 | 0 | 0 | 0 | 0 | 0 | 0 | 0 | 0 | 0 | 0 | 0 | 0 | 0 | 0 | 0 | 0 | 0 | 0 | 0 | 0 | 0 |
| rCAC | 0 | 0 | 0 | 0 | 0 | 0 | 0 | 0 | 0 | 1 | 0 | 1 | 0 | 1 | 0 | 1 | 0 | 0 | 0 | 0 | 1 | 0 | 1 | 0 | 1 | 1 | 0 | 1 | 0 | 0 | 0 | 0 | 0 | 0 | 1 | 0 | 0 | 0 | 0 | 0 | 0 | 0 | 1 | 0 | 0 | 0 | 0 | 0 | 0 | 0 | 0 | 0 | 0 | 0 | 0 | 1 | 0 | 1 | 1 | 0 | 1 | 0 | 0 | 0 | 0 | 0 |
| rCMF | 0 | 0 | 0 | 0 | 0 | 0 | 0 | 1 | 1 | 0 | 0 | 0 | 0 | 0 | 1 | 0 | 0 | 1 | 0 | 0 | 0 | 1 | 1 | 1 | 0 | 0 | 1 | 1 | 1 | 1 | 0 | 0 | 0 | 0 | 0 | 0 | 0 | 0 | 0 | 0 | 0 | 0 | 0 | 0 | 0 | 0 | 0 | 0 | 0 | 0 | 0 | 0 | 0 | 0 | 0 | 0 | 0 | 0 | 0 | 0 | 1 | 0 | 0 | 0 | 0 | 0 |
| rCUN | 0 | 0 | 0 | 0 | 0 | 0 | 1 | 1 | 0 | 1 | 1 | 0 | 1 | 0 | 1 | 0 | 0 | 0 | 0 | 0 | 1 | 0 | 0 | 0 | 1 | 0 | 0 | 0 | 1 | 1 | 0 | 0 | 0 | 0 | 0 | 0 | 1 | 0 | 0 | 0 | 0 | 0 | 1 | 1 | 0 | 1 | 0 | 0 | 0 | 0 | 0 | 0 | 0 | 1 | 0 | 0 | 0 | 1 | 0 | 0 | 0 | 1 | 0 | 0 | 0 | 0 |
| rENT | 0 | 0 | 0 | 0 | 0 | 0 | 0 | 0 | 0 | 0 | 0 | 1 | 0 | 0 | 0 | 0 | 0 | 0 | 0 | 0 | 0 | 0 | 0 | 0 | 0 | 0 | 0 | 0 | 0 | 0 | 0 | 0 | 0 | 0 | 0 | 0 | 0 | 0 | 0 | 0 | 0 | 0 | 0 | 0 | 0 | 0 | 0 | 0 | 0 | 0 | 0 | 0 | 0 | 0 | 0 | 0 | 0 | 0 | 0 | 0 | 0 | 0 | 0 | 0 | 0 | 0 |
| rFP | 0 | 0 | 0 | 0 | 0 | 0 | 0 | 0 | 0 | 0 | 0 | 1 | 0 | 1 | 0 | 0 | 0 | 0 | 0 | 0 | 0 | 0 | 0 | 0 | 0 | 1 | 1 | 1 | 0 | 1 | 0 | 1 | 0 | 0 | 0 | 0 | 0 | 0 | 1 | 0 | 0 | 0 | 0 | 0 | 1 | 0 | 1 | 0 | 0 | 0 | 0 | 0 | 0 | 0 | 0 | 0 | 0 | 0 | 0 | 0 | 1 | 0 | 0 | 0 | 0 | 0 |
| rFUS | 1 | 0 | 0 | 1 | 0 | 0 | 0 | 1 | 1 | 1 | 1 | 1 | 1 | 0 | 1 | 1 | 1 | 1 | 0 | 1 | 1 | 0 | 1 | 0 | 1 | 0 | 1 | 1 | 1 | 1 | 1 | 0 | 0 | 0 | 0 | 0 | 0 | 0 | 0 | 0 | 0 | 0 | 1 | 0 | 0 | 0 | 0 | 0 | 0 | 0 | 0 | 0 | 0 | 0 | 0 | 1 | 0 | 1 | 0 | 0 | 1 | 0 | 0 | 0 | 0 | 0 |
| rIP | 1 | 0 | 1 | 1 | 0 | 0 | 1 | 0 | 1 | 0 | 1 | 0 | 0 | 0 | 1 | 0 | 0 | 1 | 0 | 0 | 0 | 1 | 1 | 1 | 1 | 0 | 1 | 1 | 1 | 1 | 1 | 0 | 1 | 0 | 0 | 0 | 1 | 0 | 0 | 0 | 0 | 0 | 0 | 0 | 0 | 0 | 0 | 0 | 1 | 0 | 0 | 0 | 0 | 0 | 0 | 0 | 0 | 1 | 0 | 0 | 0 | 1 | 0 | 0 | 0 | 0 |
| rIT | 1 | 0 | 1 | 0 | 0 | 0 | 1 | 1 | 0 | 0 | 1 | 0 | 1 | 0 | 1 | 0 | 0 | 1 | 1 | 1 | 0 | 0 | 0 | 1 | 0 | 1 | 1 | 1 | 1 | 1 | 1 | 1 | 0 | 0 | 0 | 0 | 0 | 0 | 0 | 0 | 0 | 0 | 0 | 0 | 0 | 0 | 0 | 0 | 0 | 0 | 0 | 0 | 0 | 0 | 0 | 0 | 0 | 0 | 0 | 0 | 0 | 0 | 0 | 0 | 0 | 0 |
| rISTC | 0 | 1 | 0 | 1 | 0 | 0 | 1 | 0 | 0 | 0 | 0 | 0 | 1 | 1 | 0 | 1 | 1 | 0 | 0 | 0 | 1 | 0 | 1 | 0 | 1 | 1 | 0 | 1 | 1 | 1 | 0 | 0 | 0 | 0 | 1 | 0 | 1 | 0 | 0 | 0 | 0 | 0 | 1 | 0 | 0 | 0 | 0 | 0 | 1 | 0 | 0 | 0 | 0 | 1 | 0 | 1 | 0 | 1 | 1 | 0 | 0 | 0 | 0 | 0 | 0 | 0 |
| rLOCC | 1 | 0 | 0 | 1 | 0 | 0 | 1 | 1 | 1 | 0 | 0 | 0 | 1 | 0 | 1 | 1 | 0 | 0 | 0 | 0 | 1 | 0 | 0 | 0 | 1 | 0 | 1 | 0 | 1 | 1 | 0 | 1 | 0 | 0 | 0 | 0 | 0 | 0 | 0 | 0 | 0 | 0 | 0 | 1 | 0 | 0 | 0 | 0 | 1 | 0 | 0 | 0 | 0 | 1 | 0 | 0 | 0 | 0 | 0 | 0 | 0 | 1 | 0 | 0 | 0 | 0 |
| rLOF | 0 | 1 | 0 | 0 | 1 | 1 | 1 | 0 | 0 | 0 | 0 | 0 | 0 | 1 | 0 | 0 | 0 | 1 | 1 | 1 | 0 | 0 | 0 | 0 | 0 | 1 | 1 | 1 | 0 | 1 | 0 | 1 | 0 | 0 | 0 | 0 | 0 | 0 | 0 | 0 | 0 | 0 | 0 | 0 | 1 | 0 | 1 | 0 | 0 | 0 | 0 | 0 | 0 | 0 | 0 | 0 | 0 | 0 | 1 | 0 | 1 | 0 | 0 | 0 | 0 | 0 |
| rLING | 0 | 0 | 0 | 1 | 0 | 0 | 1 | 0 | 1 | 1 | 1 | 0 | 0 | 0 | 0 | 0 | 1 | 0 | 0 | 0 | 1 | 0 | 1 | 0 | 1 | 0 | 1 | 1 | 1 | 1 | 0 | 0 | 0 | 0 | 0 | 0 | 1 | 0 | 0 | 0 | 0 | 0 | 0 | 0 | 0 | 1 | 0 | 0 | 0 | 0 | 0 | 0 | 0 | 1 | 0 | 1 | 0 | 1 | 0 | 0 | 0 | 0 | 0 | 0 | 0 | 0 |
| rMOF | 0 | 1 | 0 | 0 | 0 | 1 | 0 | 0 | 0 | 1 | 0 | 1 | 0 | 0 | 0 | 0 | 0 | 0 | 0 | 0 | 0 | 0 | 0 | 0 | 1 | 1 | 1 | 1 | 0 | 1 | 0 | 0 | 0 | 0 | 0 | 0 | 0 | 0 | 1 | 0 | 0 | 0 | 0 | 0 | 1 | 0 | 1 | 0 | 0 | 0 | 0 | 0 | 0 | 0 | 0 | 0 | 0 | 0 | 1 | 0 | 1 | 0 | 0 | 0 | 0 | 0 |
| rMT | 1 | 0 | 1 | 1 | 0 | 0 | 1 | 1 | 1 | 0 | 1 | 0 | 0 | 0 | 0 | 0 | 0 | 1 | 1 | 1 | 0 | 0 | 0 | 1 | 1 | 1 | 1 | 1 | 1 | 1 | 1 | 1 | 0 | 0 | 0 | 0 | 0 | 0 | 0 | 0 | 0 | 0 | 0 | 0 | 0 | 0 | 0 | 0 | 0 | 0 | 0 | 0 | 0 | 0 | 0 | 0 | 0 | 0 | 0 | 0 | 0 | 0 | 0 | 0 | 0 | 0 |
| rPARC | 0 | 1 | 0 | 0 | 0 | 0 | 1 | 0 | 0 | 1 | 1 | 0 | 0 | 0 | 0 | 0 | 0 | 0 | 0 | 0 | 0 | 1 | 1 | 1 | 1 | 0 | 0 | 1 | 1 | 0 | 0 | 0 | 0 | 0 | 1 | 0 | 0 | 0 | 0 | 0 | 0 | 0 | 1 | 0 | 0 | 0 | 0 | 0 | 1 | 0 | 0 | 0 | 0 | 0 | 1 | 1 | 1 | 1 | 0 | 0 | 1 | 1 | 0 | 0 | 0 | 0 |
| rPARH | 0 | 0 | 0 | 0 | 0 | 0 | 1 | 0 | 0 | 1 | 0 | 0 | 1 | 0 | 0 | 0 | 0 | 0 | 0 | 0 | 0 | 0 | 0 | 0 | 1 | 0 | 0 | 1 | 0 | 0 | 0 | 0 | 0 | 0 | 0 | 0 | 0 | 0 | 0 | 0 | 0 | 0 | 1 | 0 | 0 | 0 | 0 | 0 | 0 | 0 | 0 | 0 | 0 | 0 | 0 | 0 | 0 | 1 | 0 | 0 | 0 | 0 | 0 | 0 | 0 | 0 |
| rPOPE | 1 | 0 | 1 | 0 | 0 | 0 | 1 | 1 | 1 | 0 | 0 | 1 | 0 | 0 | 1 | 0 | 0 | 0 | 0 | 1 | 0 | 1 | 0 | 1 | 0 | 0 | 1 | 1 | 0 | 1 | 1 | 0 | 0 | 0 | 0 | 0 | 0 | 0 | 0 | 0 | 0 | 0 | 0 | 0 | 0 | 0 | 0 | 0 | 0 | 0 | 0 | 0 | 0 | 0 | 0 | 0 | 0 | 0 | 0 | 0 | 0 | 0 | 0 | 0 | 0 | 0 |
| rPORB | 0 | 0 | 0 | 0 | 0 | 0 | 0 | 0 | 1 | 0 | 0 | 1 | 0 | 0 | 1 | 0 | 0 | 0 | 0 | 1 | 0 | 0 | 0 | 0 | 0 | 0 | 1 | 1 | 0 | 1 | 1 | 0 | 0 | 0 | 0 | 0 | 0 | 0 | 0 | 0 | 0 | 0 | 0 | 0 | 0 | 0 | 0 | 0 | 0 | 0 | 0 | 0 | 0 | 0 | 0 | 0 | 0 | 0 | 0 | 0 | 1 | 0 | 0 | 0 | 0 | 0 |
| rPTRI | 1 | 0 | 0 | 0 | 0 | 0 | 1 | 0 | 1 | 0 | 0 | 1 | 0 | 0 | 1 | 0 | 0 | 1 | 1 | 0 | 0 | 1 | 0 | 1 | 0 | 0 | 1 | 1 | 0 | 1 | 1 | 0 | 1 | 0 | 0 | 0 | 0 | 0 | 0 | 0 | 0 | 0 | 0 | 0 | 0 | 0 | 0 | 0 | 0 | 0 | 0 | 0 | 0 | 0 | 0 | 0 | 0 | 0 | 0 | 0 | 0 | 0 | 0 | 0 | 0 | 0 |
| rPCAL | 0 | 1 | 0 | 1 | 0 | 0 | 1 | 0 | 0 | 1 | 1 | 0 | 1 | 0 | 0 | 0 | 0 | 0 | 0 | 0 | 0 | 0 | 0 | 0 | 1 | 0 | 1 | 0 | 1 | 1 | 0 | 0 | 0 | 0 | 0 | 0 | 1 | 0 | 0 | 0 | 0 | 0 | 0 | 1 | 0 | 1 | 0 | 0 | 0 | 0 | 0 | 0 | 0 | 1 | 0 | 0 | 0 | 1 | 0 | 0 | 0 | 0 | 0 | 0 | 0 | 0 |
| rPSTC | 0 | 0 | 1 | 0 | 0 | 0 | 0 | 1 | 0 | 0 | 0 | 0 | 0 | 0 | 0 | 1 | 0 | 1 | 0 | 1 | 0 | 0 | 1 | 1 | 1 | 0 | 1 | 1 | 1 | 1 | 1 | 0 | 1 | 0 | 0 | 0 | 0 | 0 | 0 | 0 | 0 | 0 | 0 | 0 | 0 | 0 | 0 | 0 | 1 | 0 | 0 | 0 | 0 | 0 | 1 | 0 | 1 | 1 | 0 | 0 | 0 | 1 | 0 | 0 | 0 | 0 |
| rPC | 0 | 1 | 1 | 0 | 0 | 0 | 1 | 1 | 0 | 1 | 0 | 0 | 1 | 0 | 0 | 1 | 0 | 0 | 0 | 0 | 0 | 1 | 0 | 1 | 1 | 1 | 0 | 1 | 0 | 0 | 0 | 0 | 0 | 0 | 1 | 0 | 0 | 0 | 0 | 0 | 0 | 0 | 1 | 0 | 0 | 0 | 0 | 0 | 1 | 0 | 0 | 0 | 0 | 0 | 0 | 1 | 0 | 1 | 0 | 0 | 1 | 0 | 0 | 0 | 0 | 0 |
| rPREC | 1 | 0 | 1 | 0 | 0 | 0 | 0 | 1 | 1 | 0 | 0 | 0 | 0 | 0 | 1 | 1 | 0 | 1 | 0 | 1 | 0 | 1 | 1 | 0 | 1 | 0 | 1 | 1 | 1 | 1 | 1 | 0 | 1 | 0 | 0 | 0 | 0 | 0 | 0 | 0 | 0 | 0 | 0 | 0 | 0 | 0 | 0 | 0 | 1 | 0 | 0 | 0 | 0 | 0 | 0 | 0 | 1 | 0 | 0 | 0 | 1 | 0 | 0 | 0 | 0 | 0 |
| rPCUN | 0 | 1 | 0 | 1 | 0 | 0 | 1 | 1 | 0 | 1 | 1 | 0 | 1 | 1 | 1 | 1 | 1 | 0 | 0 | 0 | 1 | 1 | 1 | 1 | 0 | 1 | 0 | 1 | 1 | 1 | 0 | 0 | 0 | 0 | 1 | 0 | 1 | 0 | 0 | 0 | 0 | 0 | 1 | 1 | 0 | 1 | 0 | 0 | 1 | 0 | 0 | 0 | 0 | 1 | 1 | 1 | 1 | 1 | 1 | 0 | 1 | 1 | 0 | 0 | 0 | 0 |
| rRAC | 0 | 1 | 0 | 0 | 0 | 1 | 0 | 0 | 1 | 1 | 0 | 1 | 0 | 1 | 1 | 0 | 0 | 0 | 0 | 0 | 0 | 0 | 1 | 0 | 1 | 0 | 1 | 1 | 0 | 1 | 0 | 0 | 0 | 0 | 1 | 0 | 0 | 0 | 0 | 0 | 0 | 0 | 1 | 0 | 0 | 0 | 1 | 0 | 0 | 0 | 0 | 0 | 0 | 0 | 0 | 0 | 0 | 0 | 1 | 0 | 1 | 0 | 0 | 0 | 0 | 0 |
| rRMF | 0 | 0 | 1 | 0 | 0 | 1 | 1 | 1 | 1 | 0 | 1 | 1 | 1 | 1 | 1 | 0 | 0 | 1 | 1 | 1 | 1 | 1 | 0 | 1 | 0 | 1 | 0 | 1 | 1 | 1 | 1 | 0 | 0 | 0 | 0 | 0 | 0 | 0 | 1 | 0 | 0 | 0 | 0 | 0 | 1 | 0 | 1 | 0 | 0 | 0 | 0 | 0 | 0 | 0 | 0 | 0 | 0 | 0 | 0 | 0 | 1 | 0 | 0 | 0 | 0 | 0 |
| rDF | 0 | 1 | 1 | 0 | 0 | 1 | 1 | 1 | 1 | 1 | 0 | 1 | 1 | 1 | 1 | 1 | 1 | 1 | 1 | 1 | 0 | 1 | 1 | 1 | 1 | 1 | 1 | 0 | 1 | 1 | 0 | 0 | 0 | 0 | 1 | 1 | 0 | 0 | 1 | 0 | 0 | 0 | 0 | 0 | 1 | 0 | 1 | 0 | 1 | 0 | 0 | 0 | 0 | 0 | 0 | 1 | 0 | 1 | 1 | 1 | 1 | 0 | 0 | 0 | 0 | 0 |
| rSP | 0 | 0 | 1 | 1 | 0 | 0 | 1 | 1 | 1 | 1 | 1 | 0 | 1 | 0 | 1 | 1 | 0 | 0 | 0 | 0 | 1 | 1 | 0 | 1 | 1 | 0 | 1 | 1 | 0 | 1 | 1 | 0 | 1 | 0 | 0 | 0 | 1 | 0 | 0 | 0 | 0 | 0 | 1 | 1 | 0 | 0 | 0 | 0 | 1 | 0 | 0 | 0 | 0 | 1 | 1 | 1 | 1 | 1 | 0 | 0 | 0 | 1 | 0 | 0 | 0 | 0 |
| rST | 1 | 0 | 1 | 1 | 0 | 1 | 1 | 1 | 1 | 1 | 1 | 1 | 1 | 1 | 1 | 0 | 0 | 1 | 1 | 1 | 1 | 1 | 0 | 1 | 1 | 1 | 1 | 1 | 1 | 0 | 1 | 1 | 1 | 0 | 0 | 0 | 1 | 0 | 0 | 0 | 0 | 0 | 0 | 0 | 0 | 0 | 0 | 0 | 0 | 0 | 0 | 0 | 0 | 0 | 0 | 0 | 0 | 0 | 0 | 0 | 0 | 0 | 0 | 0 | 0 | 0 |
| rSMAR | 1 | 0 | 0 | 0 | 0 | 0 | 1 | 1 | 1 | 0 | 0 | 0 | 0 | 0 | 1 | 0 | 0 | 1 | 1 | 1 | 0 | 1 | 0 | 1 | 0 | 0 | 1 | 0 | 1 | 1 | 0 | 0 | 1 | 0 | 0 | 0 | 0 | 0 | 0 | 0 | 0 | 0 | 0 | 0 | 0 | 0 | 0 | 0 | 0 | 0 | 0 | 0 | 0 | 0 | 0 | 0 | 0 | 0 | 0 | 0 | 0 | 0 | 0 | 0 | 0 | 0 |
| rTP | 0 | 0 | 0 | 0 | 0 | 1 | 0 | 0 | 1 | 0 | 1 | 1 | 0 | 0 | 1 | 0 | 0 | 0 | 0 | 0 | 0 | 0 | 0 | 0 | 0 | 0 | 0 | 0 | 1 | 0 | 0 | 0 | 0 | 0 | 0 | 0 | 0 | 0 | 0 | 0 | 0 | 0 | 0 | 0 | 0 | 0 | 0 | 0 | 0 | 0 | 0 | 0 | 0 | 0 | 0 | 0 | 0 | 0 | 0 | 0 | 0 | 0 | 0 | 0 | 0 | 0 |
| rTT | 1 | 0 | 0 | 0 | 0 | 0 | 0 | 1 | 0 | 0 | 0 | 0 | 0 | 0 | 0 | 0 | 0 | 0 | 0 | 1 | 0 | 1 | 0 | 1 | 0 | 0 | 0 | 0 | 1 | 1 | 1 | 0 | 0 | 0 | 0 | 0 | 0 | 0 | 0 | 0 | 0 | 0 | 0 | 0 | 0 | 0 | 0 | 0 | 0 | 0 | 0 | 0 | 0 | 0 | 0 | 0 | 0 | 0 | 0 | 0 | 0 | 0 | 0 | 0 | 0 | 0 |
| lBSTS | 0 | 0 | 0 | 0 | 0 | 0 | 0 | 0 | 0 | 0 | 0 | 0 | 0 | 0 | 0 | 0 | 0 | 0 | 0 | 0 | 0 | 0 | 0 | 0 | 0 | 0 | 0 | 0 | 0 | 0 | 0 | 0 | 0 | 0 | 0 | 1 | 0 | 0 | 0 | 1 | 1 | 1 | 0 | 1 | 0 | 0 | 0 | 1 | 0 | 0 | 1 | 0 | 1 | 0 | 0 | 0 | 1 | 0 | 0 | 0 | 0 | 0 | 1 | 1 | 0 | 1 |
| lCAC | 0 | 1 | 0 | 0 | 0 | 0 | 0 | 0 | 0 | 1 | 0 | 0 | 0 | 0 | 0 | 1 | 0 | 0 | 0 | 0 | 0 | 0 | 1 | 0 | 1 | 1 | 0 | 1 | 0 | 0 | 0 | 0 | 0 | 0 | 0 | 0 | 1 | 0 | 0 | 0 | 0 | 0 | 1 | 0 | 0 | 1 | 1 | 0 | 1 | 1 | 0 | 0 | 0 | 1 | 0 | 1 | 0 | 1 | 1 | 0 | 1 | 0 | 0 | 0 | 0 | 0 |
| lCMF | 0 | 0 | 0 | 0 | 0 | 0 | 0 | 0 | 0 | 0 | 0 | 0 | 0 | 0 | 0 | 0 | 0 | 0 | 0 | 0 | 0 | 0 | 0 | 0 | 0 | 0 | 0 | 1 | 0 | 0 | 0 | 0 | 0 | 1 | 0 | 0 | 0 | 0 | 0 | 1 | 1 | 1 | 0 | 1 | 0 | 0 | 0 | 1 | 1 | 0 | 1 | 0 | 1 | 0 | 1 | 1 | 1 | 0 | 0 | 1 | 1 | 1 | 0 | 1 | 0 | 0 |
| lCUN | 0 | 0 | 0 | 1 | 0 | 0 | 0 | 1 | 0 | 1 | 0 | 0 | 1 | 0 | 0 | 0 | 0 | 0 | 0 | 0 | 1 | 0 | 0 | 0 | 1 | 0 | 0 | 0 | 1 | 1 | 0 | 0 | 0 | 0 | 1 | 0 | 0 | 0 | 0 | 1 | 1 | 1 | 1 | 1 | 0 | 1 | 0 | 0 | 0 | 1 | 0 | 0 | 1 | 1 | 0 | 0 | 1 | 1 | 0 | 0 | 1 | 1 | 1 | 0 | 0 | 1 |
| lENT | 0 | 0 | 0 | 0 | 0 | 0 | 0 | 0 | 0 | 0 | 0 | 0 | 0 | 0 | 0 | 0 | 0 | 0 | 0 | 0 | 0 | 0 | 0 | 0 | 0 | 0 | 0 | 0 | 0 | 0 | 0 | 0 | 0 | 0 | 0 | 0 | 0 | 0 | 0 | 0 | 0 | 0 | 0 | 0 | 0 | 0 | 0 | 0 | 0 | 0 | 0 | 0 | 0 | 0 | 0 | 0 | 0 | 0 | 0 | 0 | 0 | 0 | 0 | 0 | 0 | 0 |
| lFP | 0 | 0 | 0 | 0 | 0 | 1 | 0 | 0 | 0 | 0 | 0 | 0 | 0 | 1 | 0 | 0 | 0 | 0 | 0 | 0 | 0 | 0 | 0 | 0 | 0 | 0 | 1 | 1 | 0 | 0 | 0 | 0 | 0 | 0 | 0 | 0 | 0 | 0 | 0 | 0 | 0 | 0 | 0 | 0 | 1 | 0 | 1 | 0 | 0 | 0 | 0 | 0 | 0 | 0 | 0 | 0 | 0 | 0 | 1 | 1 | 1 | 0 | 0 | 0 | 0 | 0 |
| lFUS | 0 | 0 | 0 | 0 | 0 | 0 | 0 | 0 | 0 | 0 | 0 | 0 | 0 | 0 | 0 | 0 | 0 | 0 | 0 | 0 | 0 | 0 | 0 | 0 | 0 | 0 | 0 | 0 | 0 | 0 | 0 | 0 | 0 | 1 | 0 | 1 | 1 | 0 | 0 | 0 | 1 | 1 | 1 | 1 | 0 | 1 | 0 | 1 | 0 | 1 | 0 | 0 | 0 | 0 | 0 | 0 | 0 | 1 | 0 | 0 | 0 | 1 | 1 | 0 | 0 | 0 |
| lIP | 0 | 0 | 0 | 0 | 0 | 0 | 0 | 0 | 0 | 0 | 0 | 0 | 0 | 0 | 0 | 0 | 0 | 0 | 0 | 0 | 0 | 0 | 0 | 0 | 0 | 0 | 0 | 0 | 0 | 0 | 0 | 0 | 0 | 1 | 0 | 1 | 1 | 0 | 0 | 1 | 0 | 1 | 0 | 1 | 0 | 0 | 0 | 1 | 0 | 0 | 1 | 0 | 1 | 0 | 1 | 0 | 1 | 1 | 0 | 1 | 0 | 1 | 1 | 1 | 0 | 1 |
| lIT | 0 | 0 | 0 | 0 | 0 | 0 | 0 | 0 | 0 | 0 | 0 | 0 | 0 | 0 | 0 | 0 | 0 | 0 | 0 | 0 | 0 | 0 | 0 | 0 | 0 | 0 | 0 | 0 | 0 | 0 | 0 | 0 | 0 | 1 | 0 | 1 | 1 | 0 | 0 | 1 | 1 | 0 | 0 | 1 | 1 | 0 | 0 | 1 | 0 | 0 | 1 | 0 | 0 | 0 | 0 | 0 | 1 | 0 | 0 | 1 | 0 | 1 | 1 | 1 | 1 | 0 |
| lISTC | 0 | 1 | 0 | 1 | 0 | 0 | 1 | 0 | 0 | 1 | 0 | 0 | 0 | 0 | 0 | 1 | 1 | 0 | 0 | 0 | 0 | 0 | 1 | 0 | 1 | 1 | 0 | 0 | 1 | 0 | 0 | 0 | 0 | 0 | 1 | 0 | 1 | 0 | 0 | 1 | 0 | 0 | 0 | 0 | 0 | 1 | 1 | 0 | 1 | 1 | 0 | 0 | 0 | 1 | 0 | 1 | 0 | 1 | 1 | 0 | 1 | 1 | 0 | 0 | 0 | 0 |
| lLOCC | 0 | 0 | 0 | 1 | 0 | 0 | 0 | 0 | 0 | 0 | 1 | 0 | 0 | 0 | 0 | 0 | 0 | 0 | 0 | 0 | 1 | 0 | 0 | 0 | 1 | 0 | 0 | 0 | 1 | 0 | 0 | 0 | 0 | 1 | 0 | 1 | 1 | 0 | 0 | 1 | 1 | 1 | 0 | 0 | 0 | 1 | 0 | 1 | 0 | 0 | 0 | 1 | 1 | 1 | 0 | 0 | 0 | 1 | 0 | 1 | 1 | 1 | 1 | 0 | 0 | 0 |
| lLOF | 0 | 0 | 0 | 0 | 0 | 1 | 0 | 0 | 0 | 0 | 0 | 1 | 0 | 1 | 0 | 0 | 0 | 0 | 0 | 0 | 0 | 0 | 0 | 0 | 0 | 0 | 1 | 1 | 0 | 0 | 0 | 0 | 0 | 0 | 0 | 0 | 0 | 0 | 1 | 0 | 0 | 1 | 0 | 0 | 0 | 1 | 1 | 1 | 0 | 0 | 1 | 1 | 1 | 1 | 0 | 0 | 0 | 0 | 1 | 1 | 1 | 0 | 1 | 0 | 0 | 0 |
| lLING | 0 | 0 | 0 | 1 | 0 | 0 | 0 | 0 | 0 | 0 | 0 | 0 | 1 | 0 | 0 | 0 | 0 | 0 | 0 | 0 | 1 | 0 | 0 | 0 | 1 | 0 | 0 | 0 | 0 | 0 | 0 | 0 | 0 | 0 | 1 | 0 | 1 | 0 | 0 | 1 | 0 | 0 | 1 | 1 | 1 | 0 | 0 | 1 | 0 | 1 | 0 | 1 | 1 | 1 | 0 | 1 | 0 | 1 | 1 | 1 | 1 | 0 | 1 | 0 | 0 | 0 |
| lMOF | 0 | 0 | 0 | 0 | 0 | 1 | 0 | 0 | 0 | 0 | 0 | 1 | 0 | 1 | 0 | 0 | 0 | 0 | 0 | 0 | 0 | 0 | 0 | 0 | 0 | 1 | 1 | 1 | 0 | 0 | 0 | 0 | 0 | 0 | 1 | 0 | 0 | 0 | 1 | 0 | 0 | 0 | 1 | 0 | 1 | 0 | 0 | 0 | 0 | 0 | 0 | 0 | 0 | 0 | 0 | 0 | 0 | 1 | 1 | 1 | 1 | 0 | 1 | 0 | 0 | 0 |
| lMT | 0 | 0 | 0 | 0 | 0 | 0 | 0 | 0 | 0 | 0 | 0 | 0 | 0 | 0 | 0 | 0 | 0 | 0 | 0 | 0 | 0 | 0 | 0 | 0 | 0 | 0 | 0 | 0 | 0 | 0 | 0 | 0 | 0 | 1 | 0 | 1 | 0 | 0 | 0 | 1 | 1 | 1 | 0 | 1 | 1 | 1 | 0 | 0 | 0 | 1 | 1 | 0 | 0 | 1 | 0 | 0 | 1 | 0 | 0 | 1 | 0 | 1 | 1 | 1 | 1 | 1 |
| lPARC | 0 | 0 | 0 | 0 | 0 | 0 | 0 | 1 | 0 | 1 | 1 | 0 | 0 | 0 | 0 | 1 | 0 | 0 | 0 | 0 | 0 | 1 | 1 | 1 | 1 | 0 | 0 | 1 | 1 | 0 | 0 | 0 | 0 | 0 | 1 | 1 | 0 | 0 | 0 | 0 | 0 | 0 | 1 | 0 | 0 | 0 | 0 | 0 | 0 | 0 | 0 | 0 | 0 | 0 | 0 | 1 | 1 | 1 | 1 | 0 | 0 | 1 | 1 | 0 | 0 | 0 |
| lPARH | 0 | 0 | 0 | 0 | 0 | 0 | 0 | 0 | 0 | 0 | 0 | 0 | 0 | 0 | 0 | 0 | 0 | 0 | 0 | 0 | 0 | 0 | 0 | 0 | 0 | 0 | 0 | 0 | 0 | 0 | 0 | 0 | 0 | 0 | 1 | 0 | 1 | 0 | 0 | 1 | 0 | 0 | 1 | 0 | 0 | 1 | 0 | 1 | 0 | 0 | 0 | 0 | 0 | 0 | 0 | 0 | 0 | 1 | 0 | 0 | 1 | 1 | 1 | 0 | 0 | 0 |
| lPOPE | 0 | 0 | 0 | 0 | 0 | 0 | 0 | 0 | 0 | 0 | 0 | 0 | 0 | 0 | 0 | 0 | 0 | 0 | 0 | 0 | 0 | 0 | 0 | 0 | 0 | 0 | 0 | 0 | 0 | 0 | 0 | 0 | 0 | 1 | 0 | 1 | 0 | 0 | 0 | 0 | 1 | 1 | 0 | 0 | 1 | 0 | 0 | 1 | 0 | 0 | 0 | 1 | 1 | 0 | 1 | 0 | 1 | 0 | 0 | 1 | 1 | 1 | 1 | 1 | 0 | 0 |
| lPORB | 0 | 0 | 0 | 0 | 0 | 0 | 0 | 0 | 0 | 0 | 0 | 0 | 0 | 0 | 0 | 0 | 0 | 0 | 0 | 0 | 0 | 0 | 0 | 0 | 0 | 0 | 0 | 0 | 0 | 0 | 0 | 0 | 0 | 0 | 0 | 0 | 0 | 0 | 0 | 0 | 0 | 0 | 0 | 1 | 1 | 1 | 0 | 0 | 0 | 0 | 1 | 0 | 1 | 1 | 0 | 0 | 0 | 0 | 0 | 1 | 1 | 1 | 1 | 0 | 0 | 0 |
| lPTRI | 0 | 0 | 0 | 0 | 0 | 0 | 0 | 0 | 0 | 0 | 0 | 0 | 0 | 0 | 0 | 0 | 0 | 0 | 0 | 0 | 0 | 0 | 0 | 0 | 0 | 0 | 0 | 0 | 0 | 0 | 0 | 0 | 0 | 1 | 0 | 1 | 1 | 0 | 0 | 0 | 1 | 0 | 0 | 1 | 1 | 1 | 0 | 0 | 0 | 0 | 1 | 1 | 0 | 0 | 1 | 0 | 1 | 1 | 0 | 1 | 1 | 1 | 1 | 1 | 0 | 1 |
| lPCAL | 0 | 0 | 0 | 1 | 0 | 0 | 0 | 0 | 0 | 1 | 1 | 0 | 1 | 0 | 0 | 0 | 0 | 0 | 0 | 0 | 1 | 0 | 0 | 0 | 1 | 0 | 0 | 0 | 1 | 0 | 0 | 0 | 0 | 0 | 1 | 0 | 1 | 0 | 0 | 0 | 0 | 0 | 1 | 1 | 1 | 1 | 0 | 1 | 0 | 0 | 0 | 1 | 0 | 0 | 0 | 1 | 0 | 1 | 1 | 0 | 1 | 1 | 1 | 0 | 0 | 0 |
| lPSTC | 0 | 0 | 0 | 0 | 0 | 0 | 0 | 0 | 0 | 0 | 0 | 0 | 0 | 0 | 0 | 1 | 0 | 0 | 0 | 0 | 0 | 1 | 0 | 0 | 1 | 0 | 0 | 0 | 1 | 0 | 0 | 0 | 0 | 0 | 0 | 1 | 0 | 0 | 0 | 0 | 1 | 0 | 0 | 0 | 0 | 0 | 0 | 0 | 1 | 0 | 1 | 0 | 1 | 0 | 0 | 1 | 1 | 1 | 0 | 0 | 1 | 1 | 1 | 1 | 0 | 1 |
| lPC | 0 | 1 | 0 | 0 | 0 | 0 | 1 | 0 | 0 | 1 | 0 | 0 | 1 | 0 | 0 | 1 | 0 | 0 | 0 | 0 | 0 | 0 | 1 | 0 | 1 | 0 | 0 | 1 | 1 | 0 | 0 | 0 | 0 | 0 | 1 | 1 | 0 | 0 | 0 | 0 | 0 | 0 | 1 | 0 | 0 | 1 | 0 | 0 | 1 | 0 | 0 | 0 | 0 | 1 | 1 | 0 | 1 | 1 | 1 | 0 | 1 | 1 | 0 | 0 | 0 | 0 |
| lPREC | 0 | 0 | 0 | 0 | 0 | 0 | 0 | 0 | 0 | 0 | 0 | 0 | 0 | 0 | 0 | 1 | 0 | 0 | 0 | 0 | 0 | 1 | 0 | 1 | 1 | 0 | 0 | 0 | 1 | 0 | 0 | 0 | 0 | 1 | 0 | 1 | 1 | 0 | 0 | 0 | 1 | 1 | 0 | 0 | 0 | 0 | 0 | 1 | 1 | 0 | 1 | 0 | 1 | 0 | 1 | 1 | 0 | 1 | 0 | 1 | 1 | 1 | 1 | 1 | 0 | 1 |
| lPCUN | 0 | 1 | 0 | 1 | 0 | 0 | 1 | 1 | 0 | 1 | 0 | 0 | 1 | 0 | 0 | 1 | 1 | 0 | 0 | 0 | 1 | 1 | 1 | 0 | 1 | 0 | 0 | 1 | 1 | 0 | 0 | 0 | 0 | 0 | 1 | 0 | 1 | 0 | 0 | 1 | 1 | 0 | 1 | 1 | 0 | 1 | 1 | 0 | 1 | 1 | 0 | 0 | 1 | 1 | 1 | 1 | 1 | 0 | 1 | 0 | 1 | 1 | 1 | 0 | 0 | 0 |
| lRAC | 0 | 1 | 0 | 0 | 0 | 0 | 0 | 0 | 0 | 1 | 0 | 1 | 0 | 1 | 0 | 0 | 0 | 0 | 0 | 0 | 0 | 0 | 0 | 0 | 1 | 1 | 0 | 1 | 0 | 0 | 0 | 0 | 0 | 0 | 1 | 0 | 0 | 0 | 1 | 0 | 0 | 0 | 1 | 0 | 1 | 1 | 1 | 0 | 0 | 0 | 0 | 0 | 0 | 1 | 0 | 1 | 0 | 1 | 0 | 0 | 1 | 0 | 0 | 0 | 0 | 0 |
| lRMF | 0 | 0 | 0 | 0 | 0 | 0 | 0 | 0 | 0 | 0 | 0 | 0 | 0 | 0 | 0 | 0 | 0 | 0 | 0 | 0 | 0 | 0 | 0 | 0 | 0 | 0 | 0 | 1 | 0 | 0 | 0 | 0 | 0 | 0 | 0 | 1 | 0 | 0 | 1 | 0 | 1 | 1 | 0 | 1 | 1 | 1 | 1 | 1 | 0 | 0 | 1 | 1 | 1 | 0 | 0 | 0 | 1 | 0 | 0 | 0 | 1 | 1 | 1 | 0 | 0 | 0 |
| lDF | 0 | 1 | 1 | 0 | 0 | 1 | 1 | 0 | 0 | 0 | 0 | 1 | 0 | 1 | 0 | 1 | 0 | 0 | 1 | 0 | 0 | 0 | 1 | 1 | 1 | 1 | 1 | 1 | 0 | 0 | 0 | 0 | 0 | 0 | 1 | 1 | 1 | 0 | 1 | 0 | 0 | 0 | 1 | 1 | 1 | 1 | 1 | 0 | 1 | 1 | 1 | 1 | 1 | 1 | 1 | 1 | 1 | 1 | 1 | 1 | 0 | 1 | 1 | 0 | 0 | 0 |
| lSP | 0 | 0 | 0 | 1 | 0 | 0 | 0 | 1 | 0 | 0 | 1 | 0 | 0 | 0 | 0 | 1 | 0 | 0 | 0 | 0 | 0 | 1 | 0 | 0 | 1 | 0 | 0 | 0 | 1 | 0 | 0 | 0 | 0 | 0 | 0 | 1 | 1 | 0 | 0 | 1 | 1 | 1 | 1 | 1 | 0 | 0 | 0 | 1 | 1 | 1 | 1 | 1 | 1 | 1 | 1 | 1 | 1 | 1 | 0 | 1 | 1 | 0 | 1 | 1 | 0 | 1 |
| lST | 0 | 0 | 0 | 0 | 0 | 0 | 0 | 0 | 0 | 0 | 0 | 0 | 0 | 0 | 0 | 0 | 0 | 0 | 0 | 0 | 0 | 0 | 0 | 0 | 0 | 0 | 0 | 0 | 0 | 0 | 0 | 0 | 0 | 1 | 0 | 0 | 1 | 0 | 0 | 1 | 1 | 1 | 0 | 1 | 1 | 1 | 1 | 1 | 0 | 1 | 1 | 1 | 1 | 1 | 1 | 0 | 1 | 1 | 0 | 1 | 1 | 1 | 0 | 1 | 1 | 1 |
| lSMAR | 0 | 0 | 0 | 0 | 0 | 0 | 0 | 0 | 0 | 0 | 0 | 0 | 0 | 0 | 0 | 0 | 0 | 0 | 0 | 0 | 0 | 0 | 0 | 0 | 0 | 0 | 0 | 0 | 0 | 0 | 0 | 0 | 0 | 1 | 0 | 1 | 0 | 0 | 0 | 0 | 1 | 1 | 0 | 0 | 0 | 0 | 0 | 1 | 0 | 0 | 1 | 0 | 1 | 0 | 1 | 0 | 1 | 0 | 0 | 0 | 0 | 1 | 1 | 0 | 0 | 1 |
| lTP | 0 | 0 | 0 | 0 | 0 | 0 | 0 | 0 | 0 | 0 | 0 | 0 | 0 | 0 | 0 | 0 | 0 | 0 | 0 | 0 | 0 | 0 | 0 | 0 | 0 | 0 | 0 | 0 | 0 | 0 | 0 | 0 | 0 | 0 | 0 | 0 | 0 | 0 | 0 | 0 | 1 | 0 | 0 | 0 | 0 | 0 | 0 | 1 | 0 | 0 | 0 | 0 | 0 | 0 | 0 | 0 | 0 | 0 | 0 | 0 | 0 | 0 | 1 | 0 | 0 | 0 |
| lTT | 0 | 0 | 0 | 0 | 0 | 0 | 0 | 0 | 0 | 0 | 0 | 0 | 0 | 0 | 0 | 0 | 0 | 0 | 0 | 0 | 0 | 0 | 0 | 0 | 0 | 0 | 0 | 0 | 0 | 0 | 0 | 0 | 0 | 1 | 0 | 0 | 1 | 0 | 0 | 0 | 1 | 0 | 0 | 0 | 0 | 0 | 0 | 1 | 0 | 0 | 0 | 0 | 1 | 0 | 1 | 0 | 1 | 0 | 0 | 0 | 0 | 1 | 1 | 1 | 0 | 0 |

**Figure S2.** Human fiber-tract connectivity. 1: existing fiber-tract; 0: non-existing or non-tested fiber tract. Rows correspond to sources and columns to targets of a projection. The labels represent the following regions (r=right hemisphere, l=left hemisphere):

| | |
|---|---|
| BSTS | Bank of the superior temporal sulcus |
| CAC | Caudal anterior cingulate cortex |
| CMF | Caudal middle frontal cortex |
| CUN | Cuneus |
| ENT | Entorhinal cortex |
| FP | Frontal pole |
| FUS | Fusiform gyrus |
| IP | Inferior parietal cortex |
| IT | Inferior temporal cortex |
| ISTC | Isthmus of the cingulate cortex |
| LOCC | Lateral occipital cortex |
| LOF | Lateral orbitofrontal cortex |
| LING | Lingual gyrus |
| MOF | Medial orbitofrontal cortex |
| MT | Middle temporal cortex |
| PARC | Paracentral lobule |
| PARH | Parahippocampal cortex |
| POPE | Pars opercularis |
| PORB | Pars orbitalis |
| PTRI | Pars triangularis |
| PCAL | Pericalcarine cortex |
| PSTS | Postcentral gyrus |
| PC | Posterior cingulate cortex |
| PREC | Precentral gyrus |
| PCUN | Precuneus |
| RAC | Rostral anterior cingulate cortex |
| RMF | Rostral middle frontal cortex |
| SF | Superior frontal cortex |
| SP | Superior parietal cortex |
| ST | Superior temporal cortex |
| SMAR | Supramarginal gyrus |
| TP | Temporal pole |
| TT | Transverse temporal cortex |